\documentclass[twocolumn, aps, showpacs,amsmath,amssymb]{revtex4}
\usepackage{graphicx}
\usepackage{dcolumn}
\usepackage{bm}

\renewcommand{\vec}[1]{\mathbf{#1}}

\begin{document}
\title{Bayesian estimation of pulsar parameters from gravitational wave data}
\author{R\'ejean J. Dupuis}
 \email{rejean@caltech.edu}
\affiliation{California Institute of Technology, Pasadena, CA 91125, USA}
\affiliation{ University of Glasgow, Glasgow, G12 8QQ, UK}

\author{Graham Woan}
\affiliation{ University of Glasgow, Glasgow, G12 8QQ, UK}
\date{\today}
\begin{abstract}
We present a method of searching for, and parameterizing, signals
from known radio pulsars in data from interferometric gravitational
wave detectors. This method has  been applied to data from
the LIGO and GEO\,600 detectors to set upper limits on the
gravitational wave emission from several radio pulsars. Here we
discuss the nature of the signal and the performance of the
technique on simulated data.  We show how to perform a coherent
multiple detector analysis and give some insight in the covariance
between the signal parameters.
\end{abstract}

\maketitle
\section{\label{intro}Introduction}
Several kilometer-scale gravitational wave interferometers are now
under construction or actively collecting data with unprecedented
sensitivity~\cite{2004NIMPA.517..154A, 1997NuPhS..54..167C}. To make
the best use of these data, sophisticated analysis methods have been
developed to search for astrophysical signals that are doubtless
buried in the noise. A promising class of sources of gravitational
waves are rapidly rotating, and structurally asymmetric,  neutron
stars. Several mechanisms have been proposed that could support a
varying quadrupolar mass distribution in these neutron stars, and
the subsequent continuous emission of gravitational
radiation~\cite{2002PhRvD..66h4025C, Owen2005}.

In this paper we present a detailed end-to-end description of  an
analysis technique which we have developed to infer the parameters
of such sources using data from interferometric gravitational wave
detectors~\cite{Dupuis2004}.  Here we test the performance of the technique on 
simulated data. Whether a signal is clearly 
present or not, the method can be used to set upper limits on emission strength.  

This method was successfully applied first to GEO\,600 and LIGO data
from their first science run (S1) to set upper limits on the
strength of gravitational wave emission from pulsar
J1939+2134~\cite{2004PhRvD..69h2004A}. The search was modified and
expanded to 28 isolated pulsars using data from LIGO's second
science run (S2)~\cite{2005PhRvL..94r1103A}. This paper investigates
the methods used in these two papers and sets
performance benchmarks for the algorithm.

Searches for periodic gravitational wave signals from neutron stars
are conventionally classed as \emph{blind}, \emph{directed} or
\emph{targeted}. A search is blind if no source parameters (such as
sky position or spin evolution) are known a priori, so that the size
of the parameter space to be explored is maximal. Blind searches
represent a computationally demanding problem, and highly efficient
analysis techniques must be used~\cite{1998PhRvD..58f3001J,
2004PhRvD..70h2001K}.  Even with these methods, a fully coherent
search using many months of data is computationally intractable, and
the size of the parameter space has the effect of decreasing the
sensitivity of the search, as there is a good chance the noise will
imitate a relatively strong source somewhere in the space
\cite{1998PhRvD..57.2101B}. Directed (known location) and targeted
(known location and phase evolution) searches have smaller numbers
of unknown parameters, and vastly smaller parameter spaces, making
the detection level lower and increasing their sensitivity to
gravitational waves.

Radio pulsars are particularly interesting class of targeted sources
because i) we can monitor their rotation and make a very good guess
at the gravitational waveform they produce and ii) their locations
are known to high precision. In addition, the data pertaining to
each pulsar can be restricted to a very narrow spectral window. Real
interferometric data is usually contaminated by a large number of
instrumental spectral lines, so the effects of these can be reduced
significantly by analyzing only those narrow bands containing pulsar
data.

This paper is structured as follows: section~\ref{nature} describes
the nature of the gravitational wave signal that we are expecting
from pulsars. Section~\ref{heterodyne} describes how we filter and
greatly reduce the size of the data set using a semi-standard
heterodyne technique. In Section~\ref{bayesian} we present the
Bayesian methodology used for this analysis and describe the two
likelihood functions used in previous work. We demonstrate the
performance of the algorithm on simulated data in
Section~\ref{performance}. Section~\ref{conclusions} concludes the
paper with a brief summary.

\section{Nature of gravitational wave signal}
\label{nature} Here we summarize the form of gravitational waves
emitted from a rotating rigid triaxial pulsar, described in detail
in~\cite{1979PhRvD..20..351Z}. We take the special case of a
triaxial ellipsoid rotating about its principal axis and therefore
emitting all its gravitational radiation at twice the rotation
frequency.  A freely precessing neutron star, with its spin and
angular momentum axes non-aligned, would also emit at its rotation
frequency but is expected to be strongly damped
\cite{2002MNRAS.331..203J}. The regularity of signals from the
majority of radio pulsars suggests that most of them are not
precessing on a short timescales, if at all.

The gravitational wave amplitude from a triaxial neutron star seen
from Earth is
\begin{equation}
h_0 = \frac{16\pi^2 G}{c^4}\frac{I_{zz}f_{\rm r}^2}{r}\epsilon,
\label{h0}
\end{equation}
where $r$ is the distance to the pulsar,  $I_{zz}$ its moment of
inertia about the rotation (principal) axis, $f_{\rm r}$ the
rotation frequency of the pulsar, and  $\epsilon$ its equatorial
ellipticity, defined in terms of its principal moments of inertia as
\begin{equation}
\epsilon= \frac{I_{xx} - I_{yy}}{I_{zz}}.
\label{epsilon}
\end{equation}
The gravitational wave strain on the detector will be frequency
modulated, due to the relative motion of the Earth and the pulsar, and
amplitude modulated by the antenna pattern of the interferometer.
Following~\cite{1979PhRvD..20..351Z,1998PhRvD..58f3001J}, we can
describe this measured signal as
\begin{align}
h(t) =& \frac{1}{2}F_{+}(t;\psi)h_{0}(1 + \cos^{2}\iota)\cos\Phi(t) \nonumber \\
& + F_{\times} (t;\psi)h_{0}\cos \iota \sin \Phi(t), \label{hoft}
\end{align}
where $F_{+}$ and $F_\times$ are the antenna responses to the $+$
and $\times$ polarizations respectively, $\psi$ is the polarization
angle of the signal (determined by the position angle of the spin
axis, projected on the sky), $\iota$ is the inclination of the
pulsar with respect to the line-of-sight, and $\Phi(t)$ is the phase
of the gravitational wave signal \footnote{Authors do not agree on
whether one should work in terms of signal phase or rotational phase
in these expressions.  We use the convention
of~\protect{\cite{2004PhRvD..69h2004A}} and~\protect{\cite{2005PhRvL..94r1103A}}, so
that $\Phi=\Phi_{\rm sig}=2\Phi_{\rm rot}$. }.

We choose to time the signal phase evolution $\Phi(T)$ with respect
to the solar system barycenter (SSB), which is an inertial reference
frame, so that to third-order in barycentric time, $T$,
\begin{align}
 \Phi(T) =& \phi_{0} + 2\pi [f_{\rm s}(T - T_{0}) + \frac{1}{2}\dot{f_{\rm s}} (T -
       T_{0})^{2} \nonumber \\
       & + \frac{1}{6}\ddot{f_{\rm s}}(T - T_{0})^{3}],
\label{phioft}
\end{align}
where $\phi_0$ is the phase of the signal at a fiducial time $T_0$,
$f_{\rm s}$ is the frequency of the signal ($=2f_{\rm r}$),
$\dot{f_{\rm s}}$ is the first frequency derivative, and
$\ddot{f_{\rm s}}$ is the second frequency derivative, all at time
$T_0$. The transformation between the barycentric time~($T$) and the
topocentric time at the detector~($t$) is
\begin{align}
\label{time}
T = & t + \delta t \nonumber \\
= & t + \Delta_{\rm Roemer}+ \Delta_{\rm Shapiro} + \Delta_{\rm
Einstein} + \Delta_{\rm Binary},
\end{align}
where $\Delta_{\rm Roemer}$ is the classical Roemer delay,
$\Delta_{S\odot}$ is the Shapiro delay due to the curvature of
space-time near the Sun, $\Delta_{E\odot}$ is the Einstein delay due
to gravitational redshift and time dilation, and $\Delta_{\rm
Binary}$ contains corrections related to the pulsar's orbit, which
is zero for isolated neutron stars; see~\cite{1986ARA&A..24..537B} for
more details on these terms.  However for pulsars in binary
systems this term should include all the classical and relativistic
corrections for the shifts in the time-of-arrival of the signal due
to the motion of the source within the binary system. We will not
consider binary pulsars in this analysis, but for more details on
pulsar timing of binary systems see~\cite{1989ApJ...345..434T, WeisbergTaylor2004} and
references therein.

The second term in Equation~\ref{time}, the Roemer delay, is the
largest component (up to $\sim 8.5$~min) and due to the motion of
the Earth within the solar system. In terms of the Earth's motion it
is
\begin{equation}
\Delta_{\rm Roemer} = \frac{\vec{r}_{\rm d} \cdot \vec{k}}{c} +
\frac{(\vec{r}_{\rm d}\cdot\vec{k})^2 - |\vec{r}_{\rm d}|}{2cd},
\label{roemer_delay}
\end{equation}
where $\vec{r}_{\rm d}$ is the position of the detector with regard
to the SSB, $\vec{k}$ is a unit vector in the direction of the
neutron star, $c$ is the speed of light and $d$ is the distance from
the detector to the pulsar. In order to calculate the Roemer delay
we need accurate knowledge of the position of the Earth with regard
to the SSB.  For our barycentering software~\footnote{The timing
routines that we used are available in the LIGO Analysis Library
available at
http://www.lsc-group.phys.uwm.edu/daswg/projects/lal.html } we use
the solar system ephemerides published by the Jet Propulsion
Laboratory~\cite{JPL}.  The second term in
Equation~\ref{roemer_delay} is the timing parallax.  This takes
account of the curvature of the wavefronts emitted from the source
and is only significant for the closest sources.

The Shapiro delay $\Delta_{\rm Shapiro}$ is a relativistic
correction for the curvature of space-time near the SSB.  Since this
curvature is not negligible there will be an extra time delay in the
arrival of a signal.  In principle this delay can be as large as
120\,$\mu$s for signals passing near the edge of the Sun and
therefore becomes important for the analysis of signals from
millisecond pulsars over periods of $\sim 1$\,yr. The maximum
contribution from Jupiter however is only 200\,ns and would not
affect the sensitivity of a search.

The Einstein delay describes the combined effect of gravitational
redshift and time dilation due to the motion of the Earth. This
correction takes into account the varying gravitational potential
experienced by a clock on the Earth as it follows its elliptical
orbit around the Sun.  Again this does not significantly affect
signal searches.

\section{The complex heterodyne method}
\label{heterodyne} Current ground-based interferometric detectors
have broadband sensitivity to gravitational waves from frequencies
of several tens of hertz up to several kilohertz. As a search for a continuous wave signal involves
integrating for months or even years, the datasets involved can
become very large indeed. However, the signal we are trying to
extract in a targeted search is actually contained in only a very
narrow frequency band, so accurate knowledge of the spin parameters
of the source (from radio or X-ray observations) allows us to reduce
the size of this data set considerably.  To do this, we perform a
complex heterodyne, followed by filtering and resampling of the
data, to reduce its size by a factor of about $\sim 10^6$ without
loss of relevant information. Similar techniques have been applied
in a wide range of optical, radio and gravitational searches for
sinusoidal signals, most notably (in this
context)~\cite{1993PhRvD..47.3106N}.

We choose to perform a complex, slowly evolving, heterodyne on a
targeted source to precisely unwind the apparent phase evolution of
the source. However, for a signal from a pulsar recorded by an
interferometric detector, there is still a time varying component
remaining in the heterodyned signal from the antenna response
pattern of the interferometer.

Since the source moves through the antenna pattern on a timescale
that is much longer than the original periodic signal, after
heterodyning we can re-sample the data with a much reduced sampling
rate.  In practice the new sampling rate is determined by our wish
to monitor variations in the interferometer noise floor, which
changes on timescales of minutes, hours, and days. 
Since we keep both the real and imaginary part of the sample for each minute, our
effective bandwidth is 1/60\,Hz centered on the heterodyning
frequency which is the instantaneous frequency of the signal at the
detector.

We take the calibrated output of a gravitational wave detector to be
\begin{equation}
s(t) = h(t) + n(t),
\end{equation}
where $h(t)$ is a gravitational wave signal and $n(t)$ is noise that
is stationary over some time period but not necessarily Gaussian.
Using Equation~\ref{hoft} we recast the signal, $h(t)$, to
\begin{equation}
h(t) =A(t)e^{i\Phi(t)}  + A^*(t)e^{-i\Phi(t)}, \label{hsimple}
\end{equation}
where
\begin{equation}
A(t) = \frac{1}{4}F_{+}(t;\psi)h_{0}(1 + \cos^{2}\iota) -
\frac{i}{2} F_{\times}(t;\psi)h_{0}\cos \iota,
\end{equation}
and $A^*$ is the complex conjugate of $A$.
For a targeted pulsar we assume the frequency and frequency
derivative terms are known from electromagnetic observations so that
\begin{equation}
\phi(t) = \Phi(t+\delta t) - \phi_0
\end{equation}
can be calculated to high precision. The heterodyning step involves
multiplying the data from the interferometer by $e^{-i\phi(t)}$ to
give
\begin{align} s_{\rm het}(t) =& s(t)e^{-i\phi(t)} \nonumber \\
=& A(t)e^{i\phi_0} + A^*(t)e^{-i\phi_0-2i\phi(t)} +
n(t)e^{-i\phi(t)}. \label{hh}
\end{align}
The heterodyning process removes the rotational phase evolution from
the term in $A$, although this term will still vary over the day as
the source moves through the antenna pattern of the interferometer.
The second (upper sideband) term in $A^*$ will oscillate at nearly
twice the gravitational wave signal frequency.

We then apply a low-pass anti-aliasing filter to the heterodyned
data stream prior to averaging.  We note that, because the original
time series was real, we lose no independent information by
rejecting the upper sideband. We use a series of three third-order
Infinite Impulse Response (IIR) Butterworth filters to do this, with
frequency cutoffs that can be adjusted to the characteristics of the
data. The main requirement is
to prevent spectral disturbances from outside our final 1/60\,Hz
data band being aliased into our calculation of the averaged data.
The selection of the IIR filters and the sampling rate ultimately
depends on the opposing needs to over-resolve the timescales on
which the noise is non-stationary and for a narrow band to avoid
nearby spectral lines.

Finally we re-sample the filtered data to the post-filtering Nyquist
rate and average the results (now $s_{\rm het}^\prime$) over a
minute to form
\begin{equation}
 B_{k} = \frac{1}{M}\sum_{i=1}^{M} s_{\rm het}^\prime(t_i),
 \label{binning}
\end{equation}
where $k$ is the minute index and $M$ is the number of Nyquist
samples in 1 minute (typically $\sim 100$).

In practice there are computational advantages to performing the
heterodyning, filtering and re-sampling process described above in
two steps, starting with a fixed heterodyning frequency and a filter
that reduce the sampling rate to about 4\,Hz. A second (variable)
heterodyne can then be performed on the data to remove the Doppler
shifts due to the motion of the Earth. The advantage is that the
delay corrections between topocentric and barycentric
time-of-arrival need only be calculated 4 times, rather than (for LIGO and GEO) 16\,384
times, per second.

With the high frequency term in Equation~\ref{hh} suppressed we have
\begin{align}
 B_{k} =& \frac{1}{4}F_{+}(t_k;\psi)h_{0}(1 + \cos^{2}\iota)e^{i\phi_0} \nonumber \\
 &- \frac{i}{2} F_{\times}(t_k;\psi)h_{0}\cos \iota e^{i\phi_0} + n(t_k)^\prime,
 \label{Bk}
\end{align}
where $n(t_k)^\prime$ is the heterodyned and averaged complex noise
in bin $k$.   By the central limit theorem, we would expect the
noise, $n(t_k)^\prime$, to be well described by a Gaussian
distribution, although the width of this distribution may change
over time as the detector sensitivity evolves.

The heterodyned gravitational wave signal in this reduced data set
depends on the same four unknown parameters in Equations~\ref{hoft}
and~\ref{phioft}: $h_0$, $\psi$, $\phi_{0}$, and $\iota$, which can
be conveniently held as a vector, $\textrm{\bf a}$. We proceed in
the next section by calculating the (Bayesian) probability of the
data given these parameters and finally, through the application of
Bayes' Theorem and marginalization, we obtain posterior
probabilities for each of these parameters given the data collected.

\section{Bayesian formalism}
\label{bayesian} We take a straightforward Bayesian approach for the
following analysis and calculate the posterior probability,
$p(\textrm{\bf a}|\{B_k\},I)$, of the pulsar parameters ${\bf a}$
given the binned data, $\{B_k\}$. Bayes' Theorem tells us that
\begin{equation}
p(\textrm{\bf a}|\{B_k\}, I) = \frac{p(\textrm{\bf
a}|I)p(\{B_k\}|\textrm{\bf a},I)}{p(\{B_k\}|I)},
\end{equation}
where $\textrm{\bf a}$ is the set of parameters inferred from data
$\{B_k\}$, given our model $I$, and with likelihood
$p(\{B_k\}|\textrm{\bf a},I)$. $I$ remains constant throughout the
analysis, and will be dropped from the following expressions to
avoid clutter. It should of course be noted that all the inferences
we make from the data assume this model to be true.  Note that
    the posterior probability distribution given here assumes that a
    signal is present in the data.

Our prior beliefs in the value of the parameters are held in
$p(\textrm{\bf a})$ and we will use the least informative priors for
most of the parameters over their respective ranges. A change in
polarization angle of $\pi/2$ on the sky is equivalent to a change
of signal sign (i.e., a change in signal phase, $\phi_0$, of $\pi$),
so consistent priors are $\phi_{0}$ uniform over $[0,2\pi]$ and
$\psi$ uniform over $[-\pi/4,\pi/4]$.  The prior probability density
function for $\cos\iota$ is taken as uniform over $[-1,1]$,
corresponding to a uniform probability per unit solid angle for the
orientation of the spin axis.

The prior probability for $h_0$ is more interesting. In principle
the prior for $h_0$ should reflect all our initial beliefs on the
gravitational wave strength, $h_0$. If, for example we truly believe
that gravitational wave emission is powered by the loss of kinetic
energy from the pulsar (of known spindown rate), and that the moment
of inertia of the pulsar is reasonably well constrained, then we
should construct a prior that falls away sharply at strain levels
which are above those consistent with this spindown upper limit. Of
course we know that current detector sensitivities are insufficient
to detect such a signal, and as a result the prior would overwhelm
the broader likelihood function. We would learn nothing new from the
experiment since the posterior probability distribution function
(pdf) would largely resemble the prior pdf we chose.

At this stage in gravitational astronomy, a more useful statement
would be concerned with what the observations told us that was
\emph{independent} of spindown arguments, and therefore the prior
should reflect this greater sense of ignorance. We cannot exclude
the prior possibility of $h_0=0$, so a fully scale-invariant
Jeffreys prior ($\propto 1/h_0$) would not be appropriate. However,
we are interested in being able to set conservative upper limits on
the strength of any signals, and this argues in favour of using a
uniform prior for $h_0$.  A uniform prior favours larger values of
$h_0$ over smaller values (e.g., the prior probability for the range
0.1 to 1 is 10 times less than for 1 to 10), and represents, for
most, an acceptable state of optimistic ignorance. The resulting
upper limit for $h_0$ will therefore reflect the maximum value that
could reasonably be though of as consistent with the data and has
some additional merit because of that. In addition, a posterior
based on a uniform prior for $h_0$ can be interpreted as a
(marginal) likelihood for $h_0$ and more easily incorporated into
future analyses with other data.

For further discussion choosing priors in cases, similar to this
one, when the level of any signal may be below the sensitivity of
the experiment, see ~\cite{agostini99overcoming}. Ultimately, if
there is a strong detection the choice of the prior should not play
an important role in the results since the likelihood function would
be sufficiently strongly peaked to define the posterior. Conversely,
if no signal is present at the sensitivity level of the instrument,
the prior takes on a greater significance.

The full 4-dimensional posterior pdf contains all the information
from our analysis but is difficult to interpret directly. It is
therefore useful to reduce the dimensionality by marginalizing
(integrating) over the less interesting (nuisance) parameters. The
marginal distribution for one parameter can be viewed as a weighted
average of all the distributions of this parameter given all the
possible combinations of the other parameters.  The parameter in
which we are most interested is the gravitational wave amplitude
$h_0$, with a marginal pdf of
\begin{equation}
 p(h_0|\{B_k\}) \propto  \iiint p(\{B_k\}|\mathbf{a})p(\vec{a})\,{\rm d}\phi_0\,{\rm d}\psi\, {\rm
 d}\cos\iota,
\end{equation}
where the integral is performed numerically, over the full ranges of
the nuisance parameters.  The pdf for $h_0$ can then be normalised
trivially.

Even without a detection, placing upper limits on $h_0$ can be
physically interesting as we are essentially constraining the
equatorial ellipticity of the neutron star. We define the upper
limit of $h_0$ bounding 95\% of the cumulative probability (from
$h_0=0$) as the value $h_{95}$ that satisfies
\begin{equation}
0.95 = \int_{h_0=0}^{h_{95}}  p(h_0 | \{B_k\}) {\rm d}h_0.
\end{equation}
Note that such a limit can be placed on $h_0$ even if most of the
probability is to be found some distance from $h_0=0$ and a strong
signal is detected.

In order to calculate the likelihood function we need to have a
model of the signal in the processed data.  The model of the signal
that we are searching for in the data set, $\{B_k\}$, is obtained by
processing the original gravitational wave signal $h(t)$ in the same
way that we processed the data to give
\begin{align}
y(t_k;\textrm{{\bf a}}) = & \frac{1}{4}F_{+}(t_k;\psi)h_{0} (1 +  \cos^{2}\iota)e^{i\phi_{0}} \nonumber \\
 &- \frac{i}{2}F_{\times}(t_k;\psi) h_{0} \cos\iota e^{i\phi_{0}}.
 \label{model}
\end{align}
We
note that the model is complex and that the only time varying
component is the antenna pattern of the interferometers.   The Nyquist
frequency for this signal is well below our
reduced sampling rate of one $B_k$ per minute.  In the following two
sections we will present two different ways of evaluating the
likelihood function depending on whether the variance of the data is
known or unknown.

\subsection{Gaussian model -- known variance}
Here we give the expression for the likelihood function assuming
that we know, or can estimate accurately, the variance of the noise.
We assume that the data comprises $n$ samples of a signal, $y(t_k)$
(see Equation~\ref{model}), embedded in complex Gaussian noise
$N(0,\sigma_k)$ of zero mean and known overall variance
$\sigma_k^2$, so that
\begin{equation}
B_k = y_k + N(0, \sigma_k).
\end{equation}
If the set of $\{B_k\}$ are independent, the likelihood of the data
is simply the product of $n$ bivariate normal distributions. Note
that the distribution is bivariate because the data are now complex.
The real and imaginary parts of the $B_k$'s have independent noise
components, each with a variance $\sigma_k^2/2$. The likelihood of
the parameters is therefore
\begin{equation}
p(\{B_k\}|\textrm{\bf a}, \{\sigma_k\}) =
(\sqrt{2\pi}\sigma_k)^{-2n}
 \exp{\left(- \sum_{k=1}^{n} \frac{|B_k - y_k|^2}{2\sigma_k^2}
 \right)}.
 \label{GaussL}
\end{equation}
This Gaussian model for the likelihood was used for the first
GEO\,600 and LIGO analysis for signals from pulsar
J1939+2134~\cite{2004PhRvD..69h2004A}. For this analysis, the noise
level $\sigma_k$ was not known but was estimated for each $B_k$ from
the noise floor in a 4\,Hz band of data around the signal frequency,
assumed stationary for at least one minute.  This gives $240 \times
2$ points contributing to the estimate of the variance, making the
uncertainty in the point estimate of $\sigma_k$ small enough to be
ignored. The procedure is valid as long as the mean noise floor in
the band is representative of the noise floor at the signal
frequency, if there are no strong contaminating signals in the band
and if the noise is sufficiently stationary.  Although these
requirements were largely met for~\cite{2004PhRvD..69h2004A}, just
one millisecond pulsar was involved in the study and they cannot be
expected to be met in general.  To address this an alternative model
was developed for the S2 analysis~\cite{2005PhRvL..94r1103A}.

\subsection{Gaussian model -- unknown variance}
In the previous section we evaluated the likelihood function given
the noise level, $\sigma_k$, for each $B_k$. Generally however, the
noise level may not be known in advance or  may not be
well-estimated from the data.  Here we described the likelihood
function appropriate to this situation, which was used in the
analysis of the LIGO S2 data \cite{2005PhRvL..94r1103A}.

If $\sigma_k$ is estimated from a tighter bandwidth, or over a
shorter period, fewer data contribute and the uncertainty in its
value may be too large to use a point estimate alone. Within our
Bayesian framework the standard (and correct) approach is to treat
the noise level as another nuisance parameter and marginalise over
it, without computing a point estimate at all \cite{Bretthorst}.

We begin by calculating the likelihood of a subset of $m_j$
consecutive data points from $\{B_k\}$ which have a constant noise
level $\sigma_j$. Once we have that expression, we will calculate
the global likelihood simply using the product rule assuming that
each segment of data is independent. We will again define $n$ to be
the total number of data points $B_k$ and let $M$ be the number of
segments of data that we have assumed have the same noise level, so
that
\begin{equation}
n = \sum_{j=1}^{M} m_j .
\end{equation}
We can write the likelihood of the parameters, based on the $j$th
subset of data and marginalised over $\sigma_j$ as
\begin{align}
 p(\{B_k\}_j|\mathbf{a}) & \nonumber \propto \int_0^\infty p(\{B_k\}_j,\sigma_j|\mathbf{a})\,{\rm d}\sigma_j\\
                       &\propto \int_0^\infty p(\sigma_j|\mathbf{a})p(\{B_k\}_j|\mathbf{a},\sigma_j)
                        \,{\rm d}\sigma_j,
                       \label{bayesSigma}
\end{align}
where $p(\sigma_j|\mathbf{a})$ is the prior for the noise floor and
the likelihood
$p(\sigma_j|\mathbf{a})p(\{B_k\}_j|\mathbf{a},\sigma_j)$ is given by
Equation~\ref{GaussL}. As $\sigma_j$ is a non-zero scale parameter
we take a Jeffreys prior, uniform in $\log \sigma_j$:
\begin{equation}
 p(\sigma_j|\mathbf{a}) \propto \frac{1}{\sigma_j} \qquad (\sigma_j> 0).
 \label{priorSigma}
\end{equation}
Our final conclusions would be essentially unchanged if a uniform
prior was used instead of this Jeffreys prior~\cite{sivia96}.   Here we assume that the $\sigma_j$
associated with each subset $\{B_k\}_j$ is constant over the $m_j$
samples. In other words, we assume that the noise level of the
interferometer, in a narrow frequency band around the gravitational
wave signal, is stationary for this time. However, we allow the
noise floor to change between each subset of data  $\{B_k\}_j$. This
allows us to dynamically track the noise floor seen in real
interferometric data, which will inevitably vary on some timescale,
as the instrumental performance varies. The length of the $j$th
subset, over the which the data is assumed stationary, can also be
adjusted to reflect the known timescale of these variations.

Using Equations~\ref{GaussL},~\ref{bayesSigma}~and~\ref{priorSigma},
the likelihood of the parameters based on a subset $\{B_k\}_j$ of
constant noise $\sigma_j$ is
\begin{equation}
p(\{B_k\}_j|\mathbf{a}) \propto
\int_0^\infty\!\frac{1}{\sigma_j^{2m_j+1}}
\exp\left(\!-\!\sum_{k=k_{1(j)}}^{k_{2(j)}}\frac{|B_k-y_k|^2}{2\sigma_j^2}\right)\,{\rm
d}\sigma_j
\end{equation}
where $m_j = k_{2(j)}-k_{1(j)}+1$.  This reduces to
\begin{equation}
p(\{B_k\}_j|\mathbf{a}) \propto
\left(\sum_{k=k_{1(j)}}^{k_{2(j)}}|B_k - y_k|^2\right)^{-m_j}
\label{studentt},
\end{equation}
which is equivalent to a Student's $t$-distribution with $2m_j-1$
degrees of freedom. Recall that the likelihood derived in
Equation~\ref{studentt} is for a set of $m_j$ data points $B_k$ with
the same $\sigma_j$.  The joint likelihood of all the $M$ stretches
of data, taken as independent, is therefore
\begin{equation}
p(\{B_k\}|\mathbf{a}) \propto  \prod_j^M
\left(\sum_{k=k_{1(j)}}^{k_{2(j)}}|B_k - y_k|^2\right)^{-m_j}.
\label{StudentTL}
\end{equation}
We note that there is (of course) no explicit reference to the noise
level, but the lengths of the stationary intervals, $m_j$, can be
adjusted to reflect the performance of the detectors.

\section{Performance on simulated data}
\label{performance}
\subsection{Expected sensitivity}
The analysis described above is optimal (in a Bayesian sense) for
the data available from one or more science runs. It is however
instructive to examine the long-term performance of the method on a
large series of simulated data sets, both to confirm that the
average performance complies with our expectations and to ease
comparisons with methods based on sampling theory.

To do this, we calculated the $ h_{95}$ upper limits from 4\,000
simulated data sets, of length 10 days, varying the location of the
putative source in the sky and the location of the detector in each
set. The locations of the sources were picked randomly from a
uniform distribution over the sky, and the detector locations were
the GEO\,600 and the two LIGO sites.

\begin{figure}
\begin{center}
\includegraphics[width={8cm}]{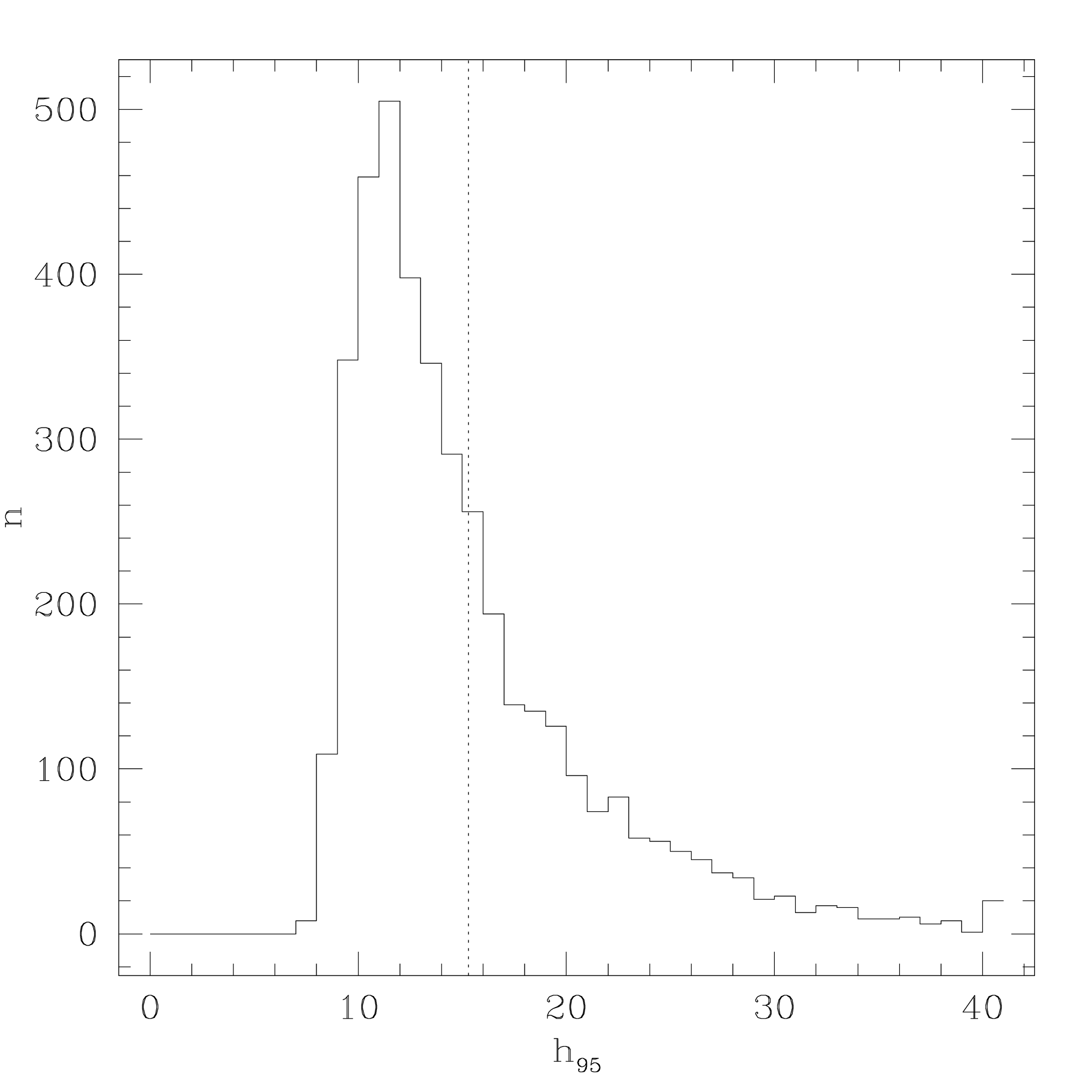}
\end{center}
\caption{ \label{sensmulti}Distribution of 95\% upper limits on
$h_0$ for 4\,000 simulations, using sources randomly located on the
sky with $S_n(f)/T = 1$.  }
\end{figure}

From these we can express the average 95\% upper limit $\langle
h_{95} \rangle$ as a function of observation time $T$ and
single-sided noise power spectral density, $S_n(f)$. Empirically,
from these simulations,
\begin{equation}
\langle h_{95} \rangle = (10.8\pm0.2)\sqrt{S_n(f) / T},
\end{equation}
where the range accounts for the location of the detector.
Figure~\ref{sensmulti} shows the distribution of $h_{95}$ that
contributed to this, for $S_n(f)/T = 1$. Note that the width and
skew of the distribution are relatively large, so the actual upper
limit from an observing run could reasonably be up to a factor of
two larger than $\langle h_{95} \rangle$.

\subsection{Combining data from a network of detectors}
\label{sec_network} Several gravitational wave detectors are
currently collecting data, and ideally we should be able to use the
observations from all detectors in a coherent manner in order to
draw the best possible inference about the source parameters.  In a
Bayesian analysis all observations enter via the likelihood
function. Assuming that the noise from each interferometer is
independent, by the product rule the global likelihood is simply the
product of the individual likelihoods.  For example, by combining
observations from GEO\,600 and the three LIGO interferometers, we
would get
\begin{equation}
p(\{B_k\}_{\rm Joint}|\mathbf{a}) = \prod_{i=1}^{4}
p(\{B_k\}_{i}|\mathbf{a},i)
\end{equation}
where the product is over the 4\,km Hanford interferometer (H1),  the
2\,km Hanford interferometer (H2), the 4\,km Livingston
interferometer (L1), and GEO\,600.

This likelihood contains all the information on the source
parameters that is contained in the data, optimally combining the
data from all the interferometers in a coherent way. Note that the
observation periods can be different and so can the sensitivity of
the detectors, although for detectors with very different
sensitivities, this will closely approximate the likelihood based on
just the most sensitive instrument.

\begin{figure}
\begin{center}
\includegraphics[width=8cm]{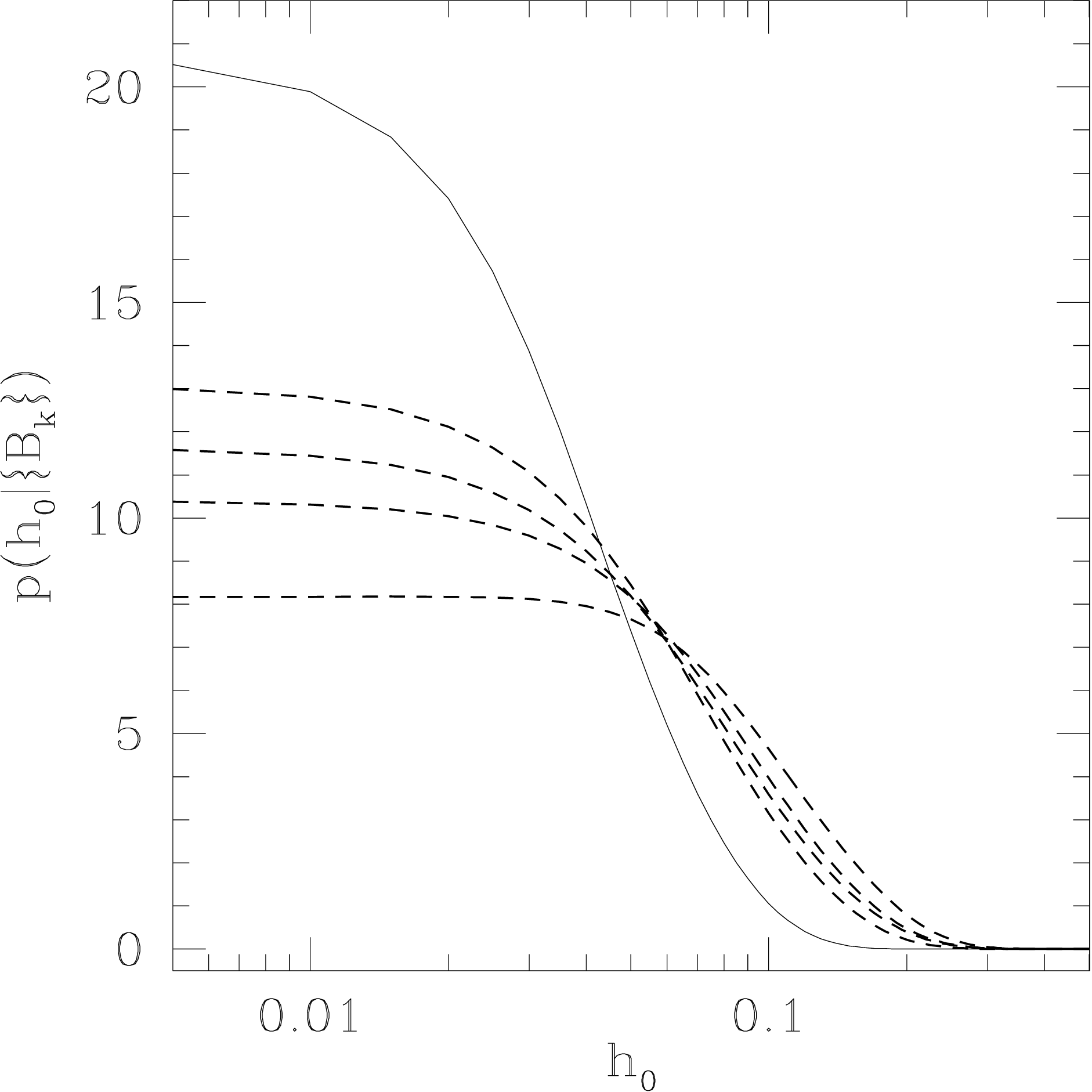}
\end{center}
\caption[]{ Multi-detector posterior pdfs
with simulated data. The solid line represents the joint
marginalised posterior pdf for $h_0$ using the data from four separate
interferometers.  The dashed lines are the corresponding pdfs from
the individual detectors.}
\label{fig_netw}
\end{figure}

For illustration, we generated four sets of data of 10 days length with
Gaussian noise ($\mu=0$ and $\sigma=1$) as if from GEO\,600 and the
three LIGO interferometers. As these IFOs are modelled as having the
same sensitivity, we would expect the coherent results to be
approximately $\sqrt{4}$ times tighter than the individual results
(distinguished from a factor of four increase in observing time only
by the differing antenna patterns of the instruments). The four
posterior pdfs for each detector as well as the joint multi-detector
posterior pdf for $h_0$ are shown in Figure~\ref{fig_netw}. The
individual 95\% upper limits are 0.15 for GEO\,600, 0.16 for H1,
0.18 for H2, and 0.13 for L1 giving an average of 0.155.  The
combined 95\% upper limit, on the other hand, is 0.08, which is
indeed approximately a factor of 2 better than the average of the
limits from the individual detectors. This technique was first
applied to real gravitational wave data in \cite{2005PhRvL..94r1103A}.
The equivalent multi-detector analysis using the ${\cal F}$-statistic method
\cite{1998PhRvD..58f3001J} has recently been developed by Cutler and
Schutz \cite{2005gr.qc.....4011C}.

It is important to realise that the posterior curve derived from a
particular observation represents a probabilistic statement about
the value of $h_0$ based on the data in hand and may, if we are
unlucky, be wildly at odds with the truth.  As a result the upper
limit derived from one instrument alone will occasionally be lower
than that from the coherent combination of instruments.

\subsection{Effects of changing the noise level}
\begin{figure}
\begin{center}
\includegraphics[width=8cm]{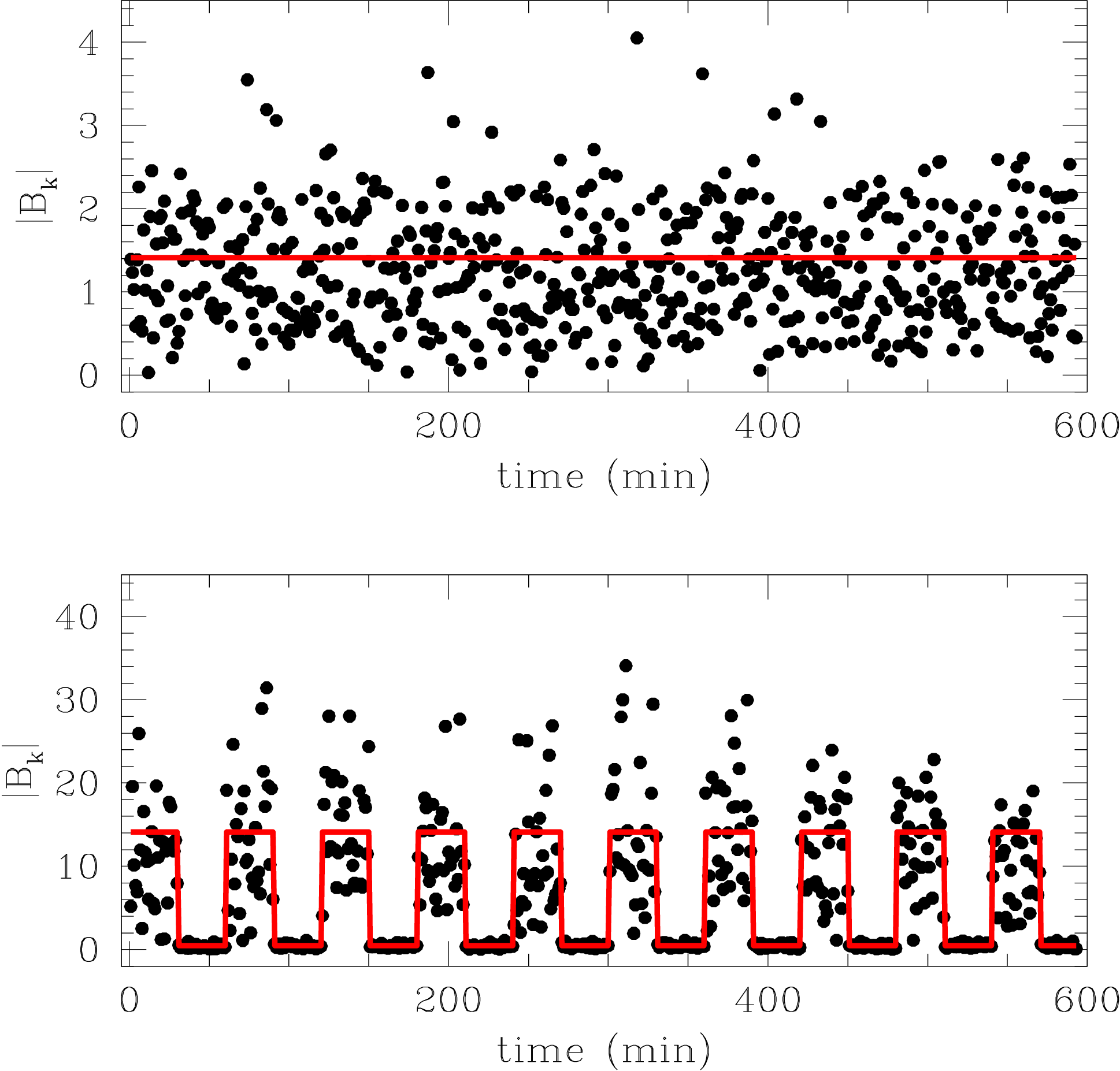}
\end{center}
\caption[]{\label{ts_sq} Time series of $B_k$ showing the effect of
a changing noise level. The dots are the $|B_k|$ and the line
represents the variance of the $B_k$'s. The top figure represents
the first data set with constant variance and the bottom figure
represents the second data set with alternating variance.}
\end{figure}

The widths of the marginal posteriors depend on both the level of
the noise and the covariance of the parameters. Here we demonstrate
the noise dependence by analysing three sets of data with different,
sometimes modulated, noise variances. Each data set corresponds to
10 days of observations. The first contains Gaussian noise with $\mu
=0$ and $\sigma =1$. For the second data set, the noise level
alternates each 30 minutes between $\sigma=10$ and
$\sigma=\frac{1}{2\sqrt{2}}$.  For the third data set, the noise
level alternates each 30 minutes between $\sigma=100$ and
$\sigma=\frac{1}{5\sqrt{2}}$.  Two time series plots showing
representative stretches of data from the first and second sets are
shown Figure~\ref{ts_sq}. 

\begin{figure}
\begin{center}
\includegraphics[width=8cm]{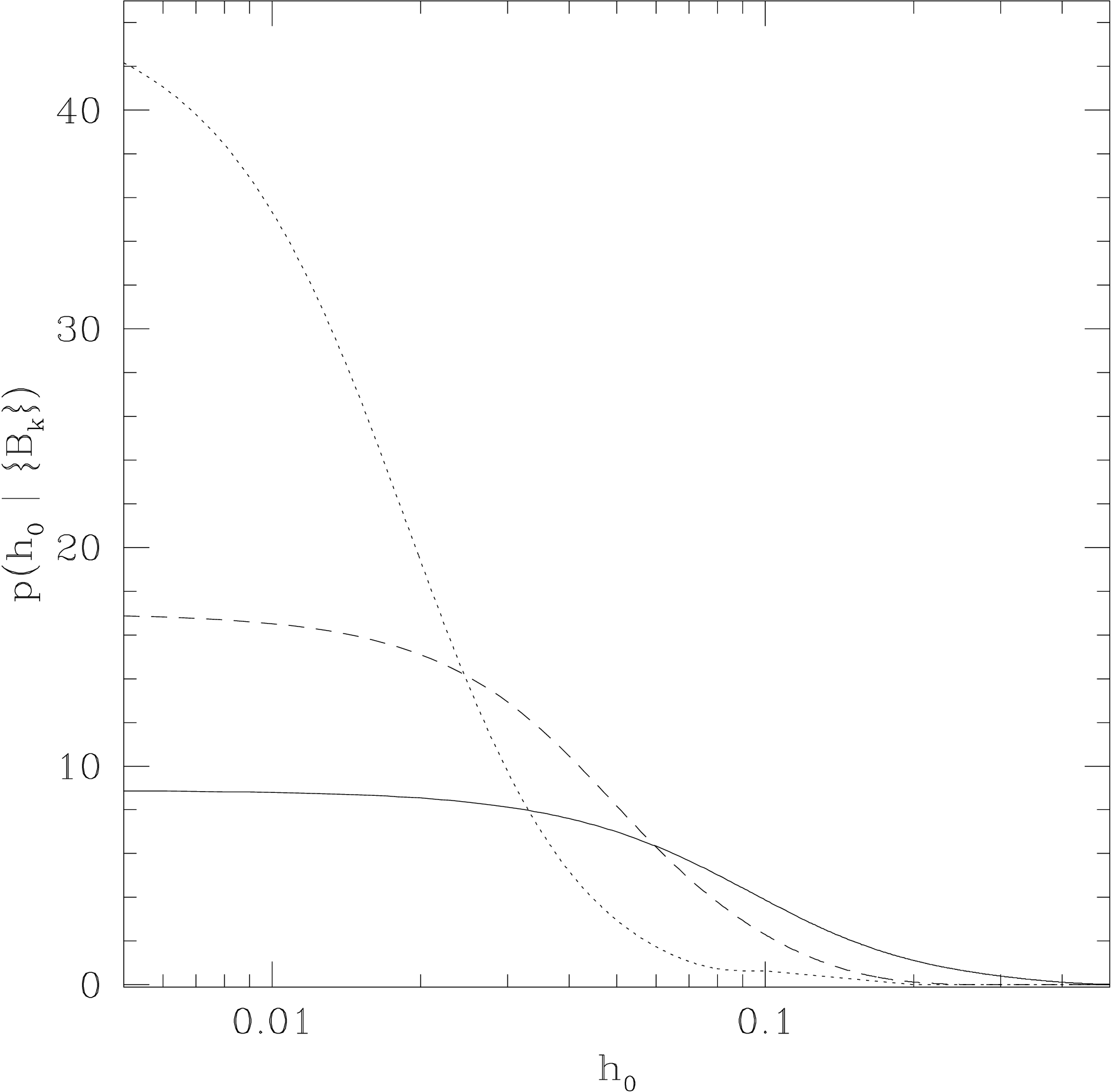}
\end{center}
\caption[]{\label{fig_sq1} Averaged marginalised posterior pdfs for
$h_0$ for three scenarios: constant unit variance (solid line),
alternating noise level between $\sigma=10$ and
$\sigma=\frac{1}{2\sqrt{2}}$ (dashed line), and alternating noise
level between $\sigma=100$ and $\sigma=\frac{1}{5\sqrt{2}}$ (dotted
line).}
\end{figure}

For this test, we repeated and averaged the posterior pdfs for 100
generations of the data sets described above. The average
marginalised posterior pdfs for $h_0$ are shown in
Figure~\ref{fig_sq1}. 
Using the 66\% upper limit on $h_0$ to characterise the width of the
pdfs we have $h_0 < 0.095$ for case 1, $h_0 < 0.047$ for case 2, and
$h_0 < 0.019$ for case 3. The second posterior is narrower than the
first by a factor of 2.02 and the third by a factor of 5.05. For no
signal, these results are roughly what we would expect: in the two
cases with alternating noise levels, about half of the data has very
low sensitivity compared to the other half, and we can assume this
noisy half does not play a significant part in the posterior.
Therefore these cases are approximately equivalent to a continuous
observation at the greater sensitivity level but for half the full
observing period, and the reduced time lowers the sensitivity by
$\sqrt{2}$. Thus compared to the first case with constant noise, we
would expect the two sets with alternating noise levels to have
widths which are narrower by factors of about 2 and 5, which is what
we see.

\subsection{Covariance between parameters}
\begin{figure}
\begin{center}
\includegraphics[width=8cm]{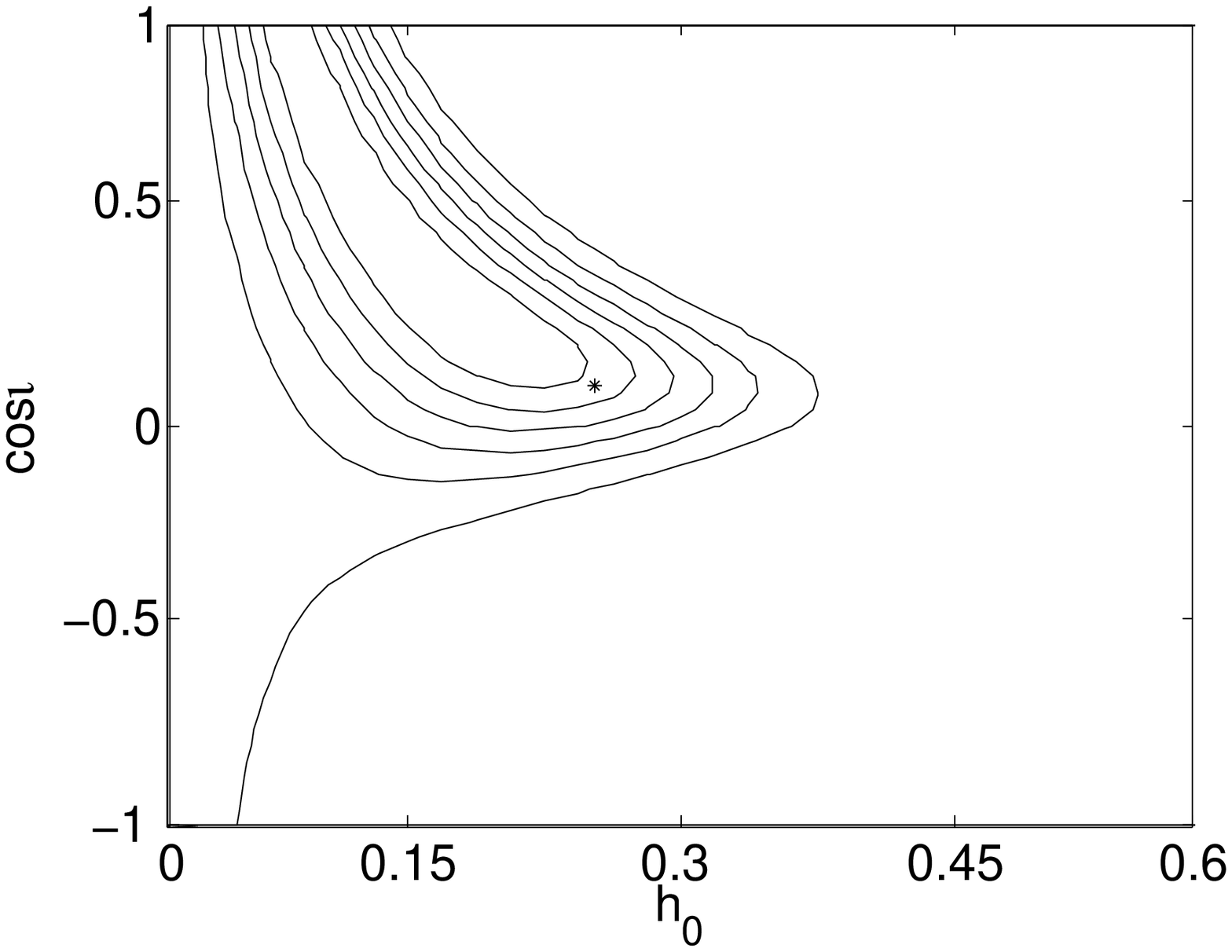}\ \ \ \
\includegraphics[width=8cm]{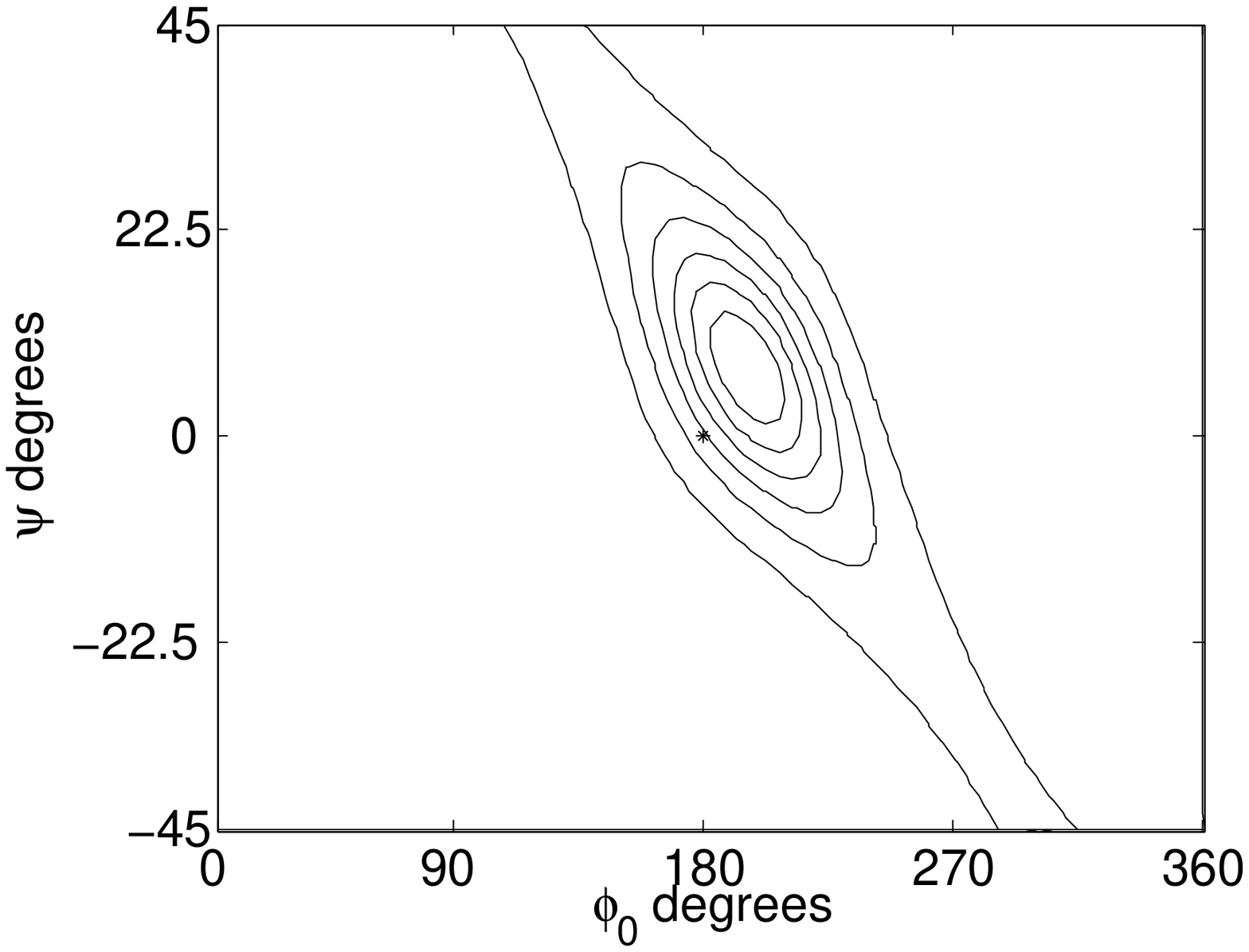}
\end{center}
\caption[]{\label{figB2} Equally-spaced contours of constant
probability density for (top) the joint posterior probability of
$\cos\iota$ and $h_0$ and (bottom) $\psi$ and $\phi_0$. The marker
indicates the location of the simulated signal. }
\end{figure}

\begin{figure}
\begin{center}
\includegraphics[width=8cm]{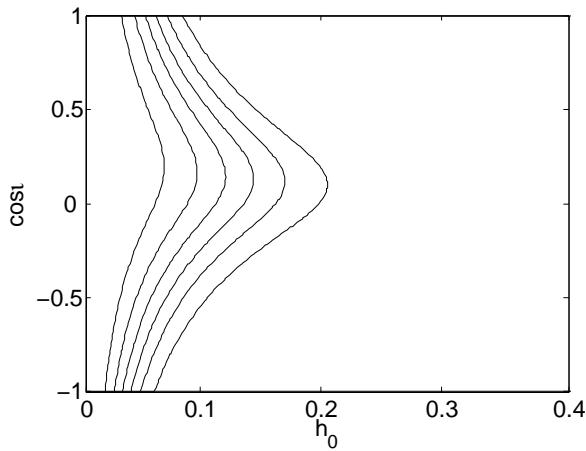}
\end{center}
\caption[]{\label{h0cosiota_nosignal} Equally-spaced contours of
constant probability, showing the covariance between $h_0$ and
$\cos\iota$, for Gaussian noise with no signal but a uniform prior
for $h_0$.}
\end{figure}

To illustrate the covariance between the signal parameters we
generated a 10 days time series containing a signal with the
following parameters and Gaussian noise of unit variance: $h_0 =
0.25$, $\cos\iota = 0.1$, $\phi_0 = 180^\circ$, $\psi = 0.0^\circ$.
It is clear from the emission model that if $\cos\iota\neq0$, $h_0$
and $\cos\iota$ are strongly anti-correlated, as are $\psi$ and
$\phi_0$. The correlation can be seen clearly in the probability
density contour plots in Figure~\ref{figB2}. 
The covariance between $\cos\iota$ and $h_0$ contributes strongly to
the overall width of the marginal pdf of $h_0$, making the precise
value of $h_0$ somewhat difficult to determine even under conditions
of relatively high signal-to-noise ratio. The correlation between
$h_0$ and $\cos\iota$ is also visible when no gravitational wave
signal is injected in the data ($h_0 = 0$), largely because of our
choice of a uniform prior for $h_0$ which holds out high hopes for a
strong signal in the data (Figure~\ref{h0cosiota_nosignal}).  When
no such signal is seen, this is interpreted as an indication that
the pulsar is oriented unfavorably and  the posterior probability
slightly increases around $\cos\iota=1$, where only the `$+$'
polarization is present.

\section{Conclusions}
\label{conclusions} In this paper we have presented an end-to-end
Bayesian method of searching for, and parameterizing, gravitational
waves from known pulsars. The method involves processing the raw
data to reduce the number of samples required for the analysis.  We
calculated the likelihood function for given model parameters from
the decimated data, so reducing computational requirements. The
algorithm has been validated by retrieving the correct signal
parameters from simulated data. We have also shown than it is easily
adapted to deal with a network  of detectors.

This methodology was initially developed for targeted searches with
known locations and spin evolutions of the sources. Further work has
now been done studying the feasibility of expanding the numbers of
parameters by taking advantage of Monte Carlo Markov Chain (MCMC)
techniques~\cite{2004PhRvD..70b2001C}. These techniques are required
when the number of unknown parameters is significantly increased,
when the method presented here would be too computationally
intensive. In a future paper we will address how this method has
been adapted to search for gravitational emission from pulsars in
binary systems. The algorithm presented in this paper, with the
binary modification, is currently being applied to GEO\,600 and LIGO
data from the S3 and S4 science runs.

\begin{acknowledgments}
The authors would like to thank the LSC Pulsar Group for useful
discussions. This work was supported by the Natural Sciences and
Engineering Research Council of Canada, Universities UK, and the
University of Glasgow. RJD would also like to acknowledge funding from the National Science Foundation.  This document has
been assigned LIGO Laboratory document number LIGO-P050046-00-Z.
\end{acknowledgments}

\bibliography{paper}
\end{document}